\documentclass[aps, prl,showpacs,preprintnumbers,amsmath,amssymb,letterpaper,showkeys]{revtex4}
\usepackage{graphicx}% Include figure files
\usepackage{dcolumn}% Align table columns on decimal point
\usepackage{bm}% bold math

\usepackage[all]{xy} % used to draw diagrams
\newcommand*{\bs}{\backslash}
\renewcommand*{\to}{\rightarrow}
\newcommand*{\from}{\leftarrow}
\newcommand{\set}[1]{ \left\{ #1 \right\} }

\newcommand*{\Cm}{$\mathrm{Ca}^{2+}$} 
\newcommand{\FIGS}{./}
\newcommand{\lfrac}[2]{\left[ #1 / #2 \right]}
\renewcommand{\sec}{s$^{-1}$}

\begin{document}
%\preprint{APS/???}

\title{Physical limits on computation by assemblies of allosteric proteins}
\author{John M. Robinson$^{1,2}$}
%\altaffiliation[Also at ]{Center for Computational Biology}%Lines break automatically or can be forced with \\
%\author{Second Author}%
% \email{Second.Author@institution.edu}
%\affiliation{Center for Computational Biology}
\affiliation{%
$^1$Department of Biochemistry and Molecular Genetics, $^2$Center for Computational Biology, University of Alabama at Birmingham, Birmingham, Alabama 35294, USA 
}%
%\altaffiliation[Also at ]{Center for Computational Biology}%Lines break automatically or can be forced with \\

%\email{jmr@uab.edu}
\email{jmr@uab.edu}

\date{\today}% It is always \today, today,
             %  but any date may be explicitly specified

\begin{abstract}
Assemblies of allosteric proteins, nano-scale Brownian computers, are the principle information processing devices in biology. The troponin C-troponin I (TnC-TnI) complex, the \Cm-sensitive regulatory switch of the heart, is a paradigm for Brownian computation.  TnC and TnI specialize in sensing (reading) and reporting (writing) tasks of computation. We have examined this complex using a newly developed phenomenological model of allostery. Nearest-neighbor-limited interactions among members of the assembly place previously unrecognized constrains the topology of the system's free energy landscape and generate degenerate transition probabilities. As a result, signaling fidelity and deactivation kinetics can not be simultaneously optimized. This trade-off places an upper limit on the rate of information processing by assemblies of allosteric proteins that couple to a  single ligand chemical bath.
\end{abstract}

% PACS, the Physics and Astronomy
                             % Classification Scheme.
                             %see http://www.aip.org/pacs/pacs06/pacs0680.html
% used -------------
%87.16.Xa Signal transduction
%87.15.He Dynamics and conformational changes -- depreciated
%87.15.km Protein-protein interactions 
%02.50.Ga Markov processes
%89.75.-k Complex systems
% unused ----------
%87.16.Ac Theory and modeling; computer simulation
%87.15.Rn Reactions and kinetics; polymerization (see also 82.39.-k Chemical
%	 kinetics in biological systems and 82.35.Pq Biopolymers,
%	 biopolymerization in physical chemistry and chemical physics)
%87.16.Yc Regulatory chemical networks

%Use showkeys class option if keyword
                              %display desired

\maketitle

%-----------------------------------------------------------------------------------------------------------------------------
\emph{Introduction---}Cells continuously regulate their function by responding to changing concentrations of diffusing molecules, called ligands.  Ligands are sensed by allosteric proteins---measuring devices that communicate ligand binding to an \emph{input domain} as a change in structure~\cite{Freire:1999fk, Hilser:2006pb} or dynamics~\cite{Kern:2003uk} in a distinct \emph{output domain}. Often, allosteric proteins are organized into allosteric supramolecular assemblies~\cite{Hartwell:1999fk, Pawson:2003oz}, where function is modularized with proteins specializing in ligand sensing or reporting. These assemblies operate as digital logic buffers, with a binary input (bound- or unbound-ligand to the input of the sensor) and binary output (bi-metastable state~\cite{Volkman:2001bs} of the reporter protein).  Important examples include the family of G-protein coupled receptors and the NF-$\kappa$B family of transcription factors~\cite{Voet:2004a}. Brownian computers are a class of computational models that use random collisions to stochastically explore the low energy portions of the computer's configuration space~\cite{Bennett:1982aa}. The allosteric supramolecular assembly is a paradigm for Brownian computation. 

Inquiry into the physical limits of computation began with the Maxwell demon (MD), a thought experiment by James Maxwell~\cite{Rex:2003ne}. The MD was the first model to include both tasks of computation: sensing and reporting, also called reading and writing.  Szilard captured the function of the MD in a computationally tractable model of an engine operating on a one dimensional gas~\cite{Szilard:1929a}. The Szillard engine drew attention to two physical concepts in computation: the relationship between information and entropy and the problem of system resetting. Both sensing and reporting require a physical memory. Landauer, representing condensed phase memory as a symmetric bi-stable potential, argued that system resetting has a minimum thermodynamic (entropic) cost of $k_B \ln 2$ dissipated per bit processed---Landauer's principle~\cite{Landauer:1961a, Bennett:1982aa}. The bi-stable potential model applies to a restricted class of simple systems, which limits the ability of Landauer's principle to reveal the physical limits of computation~\cite{Norton:2005lr}. In particular, the bi-stable potential does not adequately model the complex architecture needed to perform the sensing and reporting tasks of computation. Other attempts to place physical limits on computation~\cite{Lloyd:2000lr} recognize the role of kinetics but ignore the issue of system resetting.

Here, we explore how system complexity and system resetting impose physical limits on the rate of computation by assemblies of allosteric proteins. Our findings are based on the free energy landscape of the two-component cardiac regulatory assembly, a Brownian logic buffer. The cardiac regulatory assembly is a \Cm-sensitive switch that allows the heart to undergo periodic contraction and relaxation necessary for its pumping action~\cite{Kobayashi:2005fk}. The switch consists of the sensor protein troponin  C (TnC), a receptor for \Cm, and the reporter protein troponin I (TnI), a regulator of muscle contraction. Function is described in terms of the system's free energy landscape, resolved at the mesoscopic level. The landscape must support both activation and deactivation (resetting) components of the signaling cycle.  The topology of the landscape is constrained when communication between the system components---\Cm, sensor, and reporter---is limited to nearest-neighbor interactions. We show that in the constrainted landscape, increases in signaling fidelity are offset by decreased resetting speed. This trade-off between signaling fidelity and resetting speed can limit the rate of computation by Brownian computers. 

\emph{Signaling Dynamics---}The essential statistics and dynamics of the cardiac regulatory assembly are described by a mesoscopic phenomenological model of allostery. System dynamics involve stochastic jumps between metastable configurations of the assembly, called system-states. Each system-state $S_i$ is represented by a unique three bit binary string $(s_{0} \bs s_{1} s_{2})$.  Bit $s_{0}$ gives the liganded status of the sensor ($s_0$: 0 = unbound, 1 = bound). $s_1$ is the output state of the sensor, and $s_2$ is the output state of the reporter ($s_{i>0}$: 0 = inactive, 1 = active). The output state of the assembly is determined by the state of the reporter $s_2$. The configurational phase space, $\Omega = \set{S_i: i=1,2,\ldots, 8} = \set{(0 \bs 00), (0 \bs 10),\ldots,(1 \bs 11) }$, consists of eight system-states. Only single bit $s_j$ transitions are allowed. Transitions are represented by $\sigma_{j}^{+} \equiv (s_j: 0 \to 1)$, and $\sigma_{j}^{-} \equiv (s_j: 1 \to 0)$. In discrete time increments, $\tau \equiv t \nu_0$, where $\nu_0$ is the fastest barrier crossing attempt frequency of all system transitions (the clock frequency), the system evolves as a Markov chain
\begin{equation}
P_j(\tau+1)=\sum_{i=1}^n \Pi_{ji}(\mathbf{G}(\mu(\tau), f_i) P_i(\tau),
\label{eq:Markov_dynamics}
\end{equation}
subject to the initial distribution $\set{P_i(0)}$. $P_i(\tau)$ is the probability that the system is in state $S_i$ at  time $\tau$. System-state transitions are governed by transition probabilities $\Pi_{ji} = \Pi(S_j \from S_i)$ that depend on the free energy landscape of the system $\mathbf{G}(\mu(\tau))$, which is a function of the time-dependent chemical potential of the \Cm~ligand $\mu = k_{B}T \ln [\mathrm{Ca}^{2+}]$ and time-independent friction coefficients $f_i$ for the system-states $S_i$~\cite{Hanggi_1990a, Berezhkovskii:2005lr}. $\mu$ is relative to the standard state, taken as 1 M [\Cm]. The free energy landscape of the system is the collection of free energy surfaces  along each elementary transition. The Markov network is a graph associated with (\ref{eq:Markov_dynamics}), where the $S_i$ are represented as vertices and elementary transitions are represented as edges.

\begin{figure}[ht]
\centering
$\begin{array}{l}
\multicolumn{1}{l}{\mbox{\bf (a)}} \\ [-0.53cm] 
\includegraphics[width=7cm]{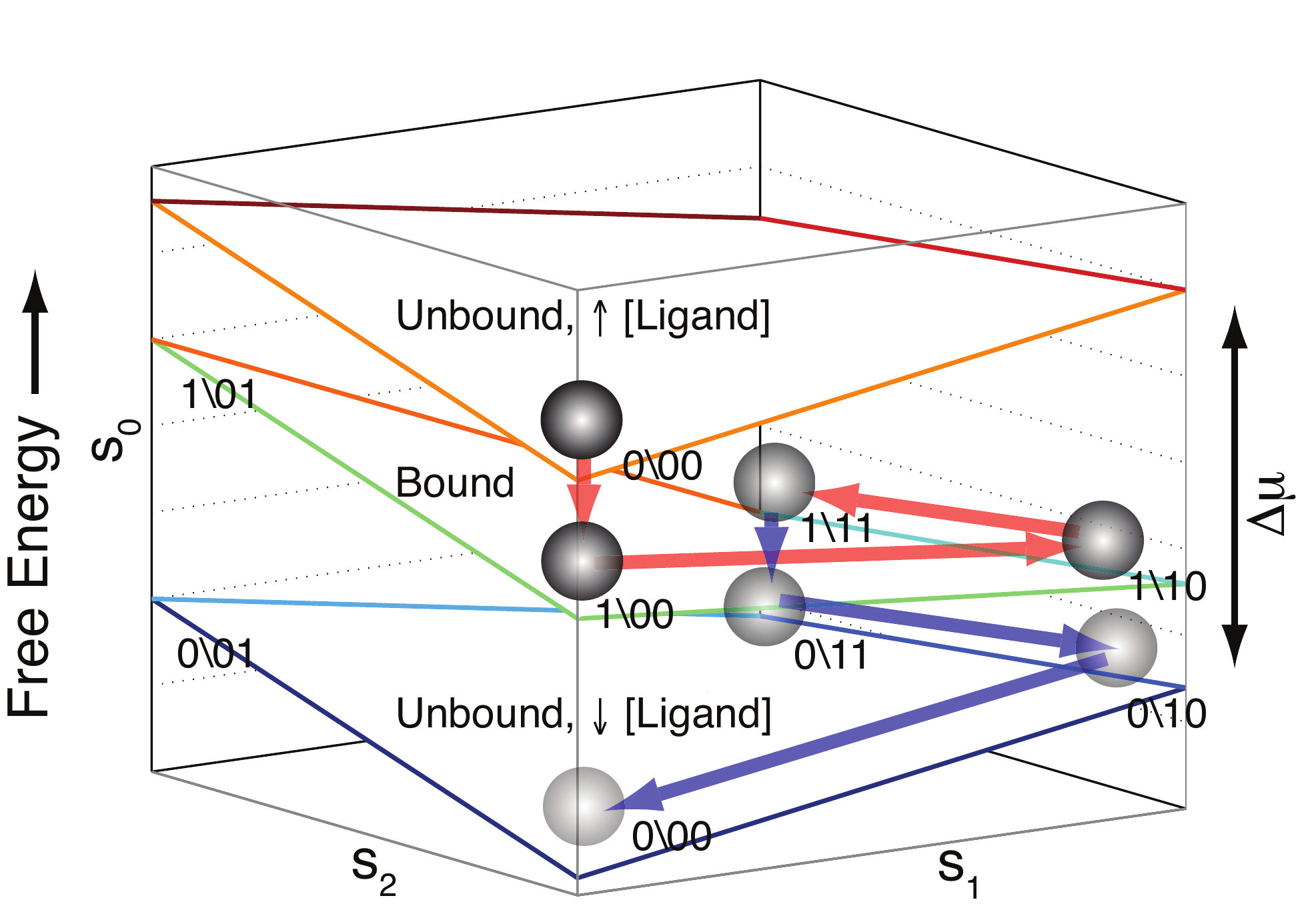} \\
\multicolumn{1}{l}{\mbox{\bf (b)}} \\ %[-0.53cm] 
\includegraphics[width=7cm]{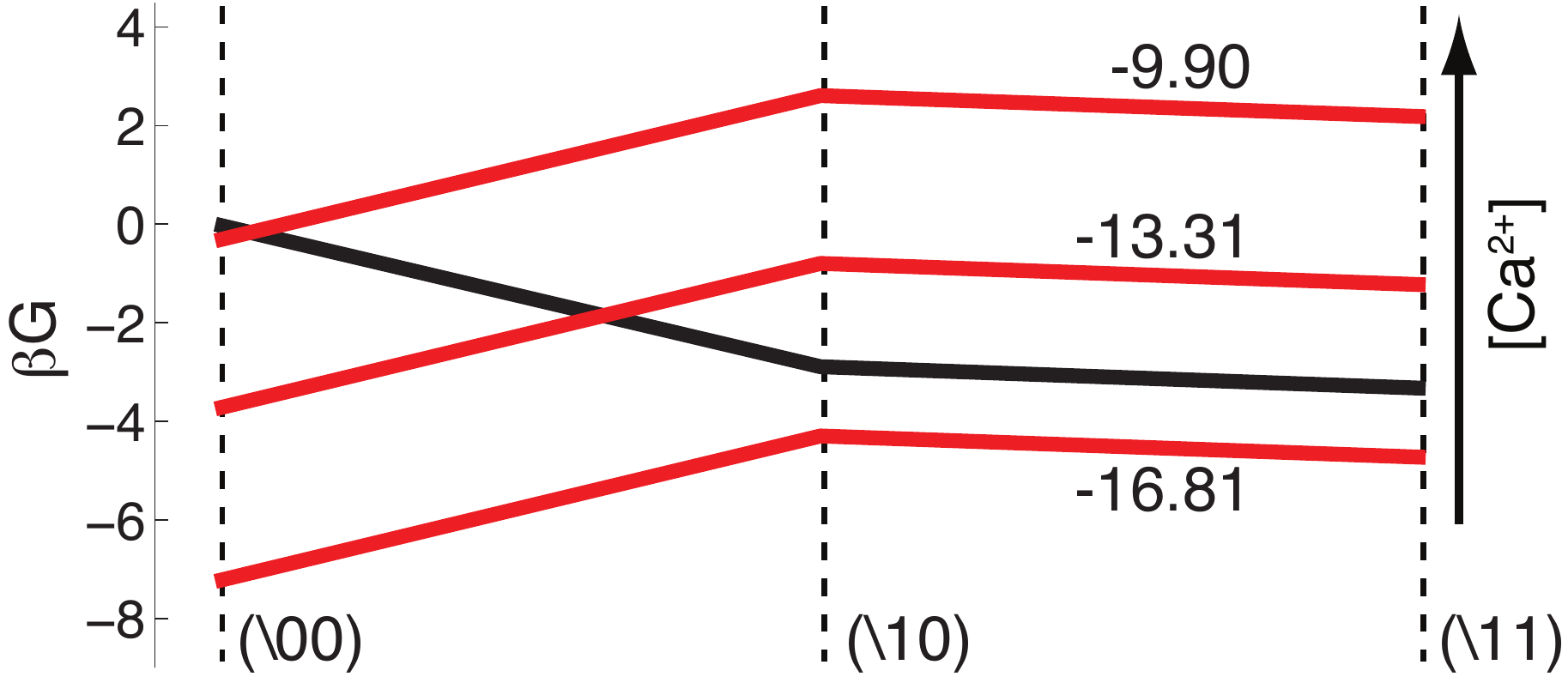}
\end{array}$
\caption
{\label{fig:landscape}
Basin-limited portions of the free energy landscape $\set{G(S_i, \mu)}$ of a two-component allosteric assembly (vertices). (a) System satisfying the 5 topological considerations (see text). Dominant pathways for activation (red arrows) and deactivation (blue arrows). (b) $\set{G(S_i, \mu)}$ of the cardiac regulatory switch shown for desaturating \Cm~($\mu = -16.81 k_{B}T$), midpoint of activation ($\mu = -13.31 k_{B}T$) and saturating \Cm~($\mu = -9.90 k_{B}T$). $G(1 \bs s_1 s_2)$ (\Cm-bound species, black) are stationary; $G(0 \bs s_1 s_2, \mu)$  (\Cm-unbound species, red) depend linearly on $\mu = k_{B}T \ln[\mathrm{Ca}^{2+}]$.
}
\end{figure}

Fig.~\ref{fig:landscape}a shows the basin-limited free energy landscape---a simplified representation of the free energy landscape---of a generic two-component allosteric assembly.  The free energies of the system-states $\set{G(S_i)}$ are shown on the z-axis of the Markov network with transitions in $s_2$, $s_1$ and $s_0$ along the x, y and z axes. By partitioning the $\set{ G(S_i) }$ into the ligand-unbound $\set{G(0 \bs s_1 s_2)}$ and ligand-bound $\set{G(1 \bs s_1 s_2)}$ surfaces, the basin-limited landscape provides a visual description of how the system's energetics supports signaling. Activation and deactivation are initiated by $\mu$-induced raising and lowering, respectively, of the $\set{G(0 \bs s_1 s_2)}$ relative to the fixed $\set{G(1 \bs s_1 s_2)}$. The $\mu$-induced change in the energy landscape cause the time-dependent population changes in (\ref{eq:Markov_dynamics}) responsible for the dissipative work of allosteric signaling. Population flux is represented as a ball rolling on the landscape. 

% ---------------------------- figure --------------------------

The topology of the basin-limited landscape $\set{ G(S_i) }$ is constrained by five considerations. 
\emph{(i)} The  free eneries of all ligand-unbound species $(0 \bs s_1 s_2)$ depend linearly on $\mu$: $G(0 \bs s_1 s_2) = G^o(0 \bs s_1 s_2) + n \mu$. $n = 1$ because cardiac troponin C has one regulatory binding site for \Cm. The free energies of the ligand bound species $G(1 \bs s_1 s_2)$ are fixed ($n = 0$) becuse the \Cm~binding site on troponin C is occupied. 
\emph{(ii)} The free energy changes along a closed path sum to zero, $\Delta G_{i \from k} + \cdots +   \Delta G_{j \from i} = 0$, where $\Delta G_{i \from k} = G(S_i) - G(S_k)$.  Free energy conservation, along with microscopic detailed balance: $\Pi_{ji} p_i = \Pi_{ij} p_j$, produces macroscopically balanced transition probabilities,
$
%\frac{ \Pi_{j \from i} }{ \Pi_{j \to i} } = \exp ( -\beta  G_{j \from i} )
\Pi_{ik} \cdots \Pi_{ji} =  \Pi_{ik} \cdots \Pi_{ji}
$, where $p_i = P^{\mathrm{eq}}_i = P_{i}(\tau \to \infty)$. \emph{(iii)} Long distance allosteric signalling is exponentially suppressed beyond the localization length of the low frequency modes that produce allostery~\cite{Bahar:1998a}. We assume that each low frequency mode is spatially limited to an assembly component and its nearest-neighbors. The assumption of nearest-neighbor-limited influence, which parallels the Markov assumption, causes certain transition probabilities to be degenerate
\begin{equation}
\Pi( \sigma_{j}^{\pm} | s_{0} \ldots s_{j-1} s_{j+1} \ldots s_{m} ) = \Pi ( \sigma_{j}^{\pm} | s_{j-1} s_{j+1} )
\label{eq:nn_influence}
\end{equation}
($m = 2$). The system-state $S_i = (s_0 \bs s_1 s_2)$ has been rewritten as $S_i = (s_j | s_{k \neq j})$ to emphasize the transition of component $j = 0,1,2$ against the fixed components ${k \neq j}$. Eq. (\ref{eq:nn_influence})  implies, for example, that the reporter $s_2$ has no direct knowledge of whether ligand $s_0$ is bound to the sensor; rather, ligand binding is communicated to the reporter by allosteric change $\sigma_1^+$ in the sensor. Degenerate transition probabilities occur for activation/deactivation of the reporter $\sigma_2^{\pm}$, $\Pi ( 0 \bs 1  \sigma_{2}^{\pm}) = \Pi ( 1 \bs 1  \sigma_{2}^{\pm})$, and for ligand binding/release $\sigma_0^{\pm}$, $\Pi ( \sigma_{0}^{\pm} \bs s_1 0) = \Pi ( \sigma_{0}^{\pm} \bs s_1 1)$. For these transitions, degeneracy is due to a degenerate free energy landscape and degenerate frictional coefficients upon which the $\Pi_{j \from i}$ depend. Particularly significant is degeneracy in the free energy change that governs switching in the reporter $\sigma_2^{\pm}$, $\Delta G( 0 \bs 1  \sigma_{2}^{\pm}) = \Delta G ( 1 \bs 1  \sigma_{2}^{\pm})$. Fig.~\ref{fig:landscape} shows how nearest-neighbor-limited influence imposes parallelpipid geometry for $\sigma_2^{\pm}$ in the free energy landscape. Parallelpipid geometry is \emph{not} imposed for $\sigma_1^{\pm}$. \emph{(iv)} Deactivation of the sensor while the reporter is active causes steric conflict, causing $G(s_0 \bs 01)$ to be very high. The high energy states $(s_0 \bs 01)$ are unvisited and can be ignored. \emph{(v)} When ligand is bound, sensor activation is favorable $\Delta G(1 \bs \sigma_1^+ 0)  < 0 $; when ligand is not bound, sensor activation is unfavorable $\Delta G(0 \bs \sigma_1^+ 0)  >  0$.  

% ------------------------------------------------------------------------------------------------------------------------------
% ------------------------------------------------------------------------------------------------------------------------------
% ------------------------------------------------------------------------------------------------------------------------------
%\section{Results}

\emph{Signaling error}---We have used time-resolved and stopped-flow F\"orster resonance energy transfer (FRET) measurements~\cite{Robinson_2007b} to parametrize the $\set{ G(S_i) }$ of the cardiac regulatory switch: $G(0 \bs 00, 0 \bs 10, 0 \bs 11) = \set{9.56, 12.52, 12.08} k_{B} T +  \mu$; $G(1 \bs 00, 1 \bs 10, 1 \bs 11) = \set{0, -2.89, -3.33 } k_{B} T$; $T = 15 $~C. All $G(S_i)$ are relative to $G(1 \bs 00)$, which is set to zero. Fig.~\ref{fig:landscape}b shows the parametrized landscape. Significantly, the free energy change that governs activation of the reporter is very small, $\Delta G(s_0 \bs 1\sigma^{+}) = -0.44 k_{B} T$.

For the assembly equilibrated under the \Cm~chemical potential $\mu$, the system-state probabilities $p_i(\mu)$ are Boltzmann distributed
\begin{equation}
p_i(\mu) = {e^{-\beta G(S_i; \mu) }} / {\sum_{\Omega} e^{-\beta G(S_i; \mu)}}.
\label{eq:Boltz}
\end{equation}
\Cm-induced signaling $s_0 \to s_2$ can be quantitated as the relative entropy change of the reporter $s_2$ in response to a change in the \Cm~chemical potential $\mu$~\cite{Cover_2006a}
\begin{equation}
\mathcal{D} \left[ \set{p^*_{s_2}}  || \set{p_{s_2}}  \right] (\mu) \equiv \sum_{ \set{s_2}} p^*_{s_2} \ln \lfrac{p^*_{s_2}}{ p_{s_2} },
\label{eq:KL}
\end{equation}
where $ p_{s_2} = \sum_{\set{s_0,s_1}} p_{s_0 \bs s_1 s_2}$, $s_i = 0,1$ are marginalized system-state probablities. The reference distribution $p^{*}_{s_2} = p_{s_2} (\mu \to - \infty)$ is the equilibrium distribution in the absence of ligand. Systems with low signaling error mirror the input at the output ($s_2 = s_0$) at all $\mu$. There are two sources of signaling error: activity in the absence of bound ligand (constitutive activation) and inactivity when ligand is bound (incomplete activation). $\mathcal{D} \left[ p^{*}_{s_2}  | p_{s_2}  \right] $ addresses only error from incomplete activation. Both sources of signaling error $\mathcal{E}(\mu)$ are treated by the conditional entropy $H[s_2 | s_0] (\mu) $ of $s_2$ given $s_0$~\cite{Cover_2006a},
\begin{equation}
\mathcal{E} \equiv H[s_2 | s_0] (\mu) = -\sum_{\set{s_0,s_2}} p_{s_0,s_2} \log_2 \lfrac{p_{s_0,s_2} }{p_{s_0} },
\label{eq:cond_ent}
\end{equation}
where $p_{s_0, s_2} = \sum_{\set{s_1}} p_{s_0 \bs s_1 s_2} $ and $ p_{s_0} = \sum_{\set{s_1,s_2}} p_{s_0 \bs s_1 s_2} $. Signaling error, $0 \leq \mathcal{E}(\mu) \leq H[s_2] \leq 1$, is a bounded logarithmic measure of the error in transmitting $s_0 \to s_2$ for the cardiac regulatory assembly in equilibrium with an externally regulated $\mu$. $\mathcal{E}_0 = H[s_2 | s_0] (-\infty) = H[s_2 | s_0 = 0]$ is the error of transmitting $s_0 = 0$. $\mathcal{E}_1 = H[s_2 | s_0] (\infty) = H[s_2 | s_0 = 1]$ is the error of transmitting $s_0 = 1$. The mean signaling error $ \bar{\mathcal{E}} \equiv \lim_{\lambda \to \infty} \lfrac{ \int_{-\lambda}^{\lambda} d \mu H[s_2 | s_0] } { \int_{-\lambda}^{\lambda} d \mu } $, which simplifies to $\bar{\mathcal{E}} = \left ( \mathcal{E}_0+\mathcal{E}_1 \right ) / 2 $, is a measure of system performance over all ligand concentrations. The fidelity of transmitting $s_0 = i$ is $\mathcal{F}_i = 1 - \mathcal{E}_i$. Overall system fidelity is $\mathcal{F} = \mathcal{F}_0 \mathcal{F}_1$.

\begin{figure}
%\centering
\includegraphics[width=8.0cm]{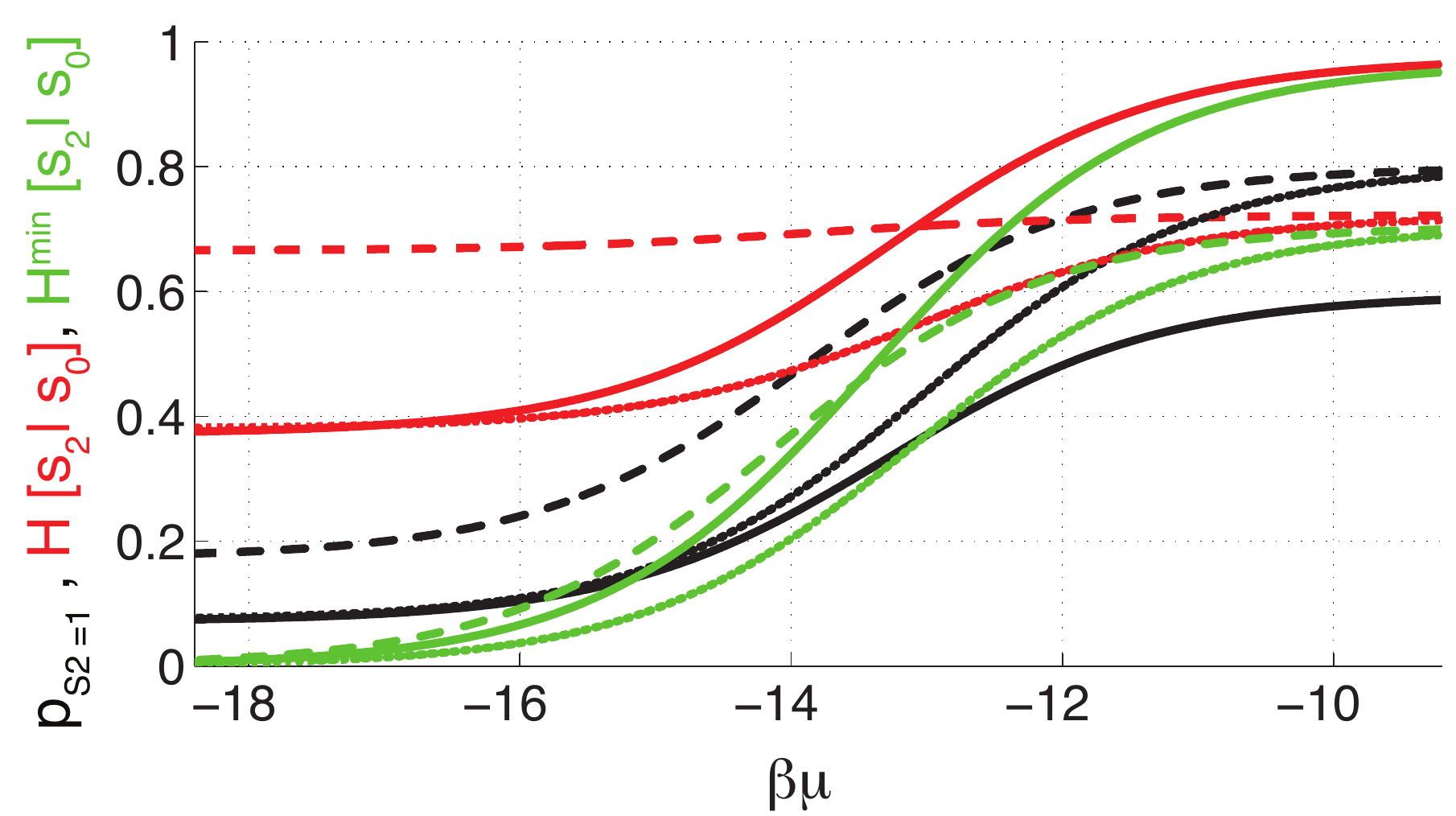}
\caption
{\label{fig:info} Signaling fidelity. Probability of being active $p_{s_2=1}$ (\ref{eq:Boltz}) (black), signaling error $\mathcal{E} = H\left[s_2 | s_0 \right]$ (\ref{eq:cond_ent}) (red), and minimum signaling error $\mathcal{E}^{\min} $ (green) vs. $\beta \mu = \ln{[\mathrm{Ca}^{2+}]}$. Experimentally measured landscape (solid lines), artificial 1 $k_B T$ decrease in $G(s_0 \bs 11)$ (dashed lines), and \emph{additional} 1 $k_B T$ decrease in $G(0 \bs 00)$ (dotted lines). 
}
\end{figure}

The probability of being active $p_{s_2=1}$ and signaling error $\mathcal{E}$ were calculated as a function of $\mu$ from the parametrized $\set{G(S_i)}$ of the cardiac regulatory assembly. They are shown as solid lines in Fig.~\ref{fig:info}. The system has low constitutive activity, $p_{s_2=1}(-\infty) = 0.07$, and incompletely activates (at high [\Cm]), $p_{s_2=1}(\infty)= 0.60$. This generates reasonably low error at low [\Cm], $\mathcal{E}_0 = 0.37$, but causes substantial error at high [\Cm], $\mathcal{E}_1 = 0.97$. The average signaling error is large, $\bar{\mathcal{ E }} = 0.67$. On paper, one can attempt to improve $\bar{\mathcal{ E }} $ subject to the constraint that $\Delta G(0 \bs 1\sigma_{2}^{-}) = \Delta G(1 \bs 1\sigma_{2}^{-})$, abbreviated hence as $ \Delta G$. The dashed lines in Fig.~\ref{fig:info} show   $p_{s_2=1}$ and $\mathcal{E}$ after $G(0 \bs 11)$ and $G(1 \bs 11)$ have been jointly decreased by 1 $k_B T$. The modified landscape raises the maximum ligand-induced activity, $p_{s_2=1}(\infty) = 0.80$ but also increases constituitive activity, $p_{s_2=1}(-\infty) = 0.17$. The desired reduction in $\mathcal{E}_1= 0.72$ is offset by an increased $\mathcal{E}_0= 0.66$. In the re-designed system, $\bar{\mathcal{ E }} = 0.69$ actually increases. Recall that nearest-neighbor-limited influence does not constrain the free energy changes for the sensor, so $G(0 \bs 00)$ and $G(1 \bs 00)$ can be independently varied. The dotted lines in Fig.~\ref{fig:info} show $p_{s_2=1}$ and $\mathcal{E}$ after an additional adjustment to the landscape---lowering $G(0 \bs 00)$ by 1 $k_B T$. The adjustment restores low constituitive activity, $p_{s_2=1}(-\infty) = 0.07$ and retains the gains in maximal activity, $p_{s_2=1}(\infty) = 0.80$. $\mathcal{E}_0 = 0.38$, $\mathcal{E}_1= 0.72$, and $\bar{\mathcal{E}}= 0.55$. The modified landscape reduces average signaling error $\bar{\mathcal{E}}$ by 18\%. Below, we show that these gains in signaling fidelity are offset by a reduced rate of deactivation. 

What is the minimum possible signaling error $\bar{\mathcal{ E }}^{\min}$ for a given $\Delta G$? An error-minimizing landscape with fixed $\Delta G$ populates $(1 \bs 10)$ and $(1 \bs 11)$ when \Cm~is bound and populates only $(0 \bs 00)$ when \Cm~is unbound (i.e. $\mathcal{E}_0^{\min} = 0$). For all other species $S_i$, $p_i = 0$. The modified energy landscape of the cardiac regulatory assembly that produces $\bar{\mathcal{ E }}^{\min}$ is $\beta G(0 \bs 00, 0 \bs 10, 0 \bs 11) = \set{-2.96, \alpha, \alpha-\Delta G } + 15.40 + \beta \mu$; and $\beta G(1 \bs 00, 1 \bs 10, 1 \bs 11) = \set{\alpha+2.89, 0,  -\Delta G  }$, where $\alpha \to \infty$. The error-minimizing landscape was subjected the same perturbations as above---1 $k_B T$ decrease in $G(s_0 \bs 11)$ and an additional 1 $k_B T$ decrease in $G(0 \bs 00)$. The green lines in Fig.~\ref{fig:info} show $\mathcal{E}^{\min}$ in the three landscapes.  As required, $\mathcal{E} \geq \mathcal{E}^{\min}$ for all $\mu$. Also, $\mathcal{E}_0^{\min} = 0$ and $\mathcal{E}_1^{\min} \to \mathcal{E}_1$. From (\ref{eq:Boltz}) and (\ref{eq:cond_ent}),
\begin{equation}
\mathcal{E}_1^{\min} = \log_{2} \left( 1 + e^{- \beta \Delta G} \right) + \beta \Delta G \frac{  e^{- \beta \Delta G}}{1 + e^{- \beta \Delta G}},
\label{eq:signaling_error}
\end{equation}
is the minimum attainable signaling error for a given $\Delta G \geq 0$, and $\mathcal{ F }^{\max}=  ( 1- \mathcal{E}_0^{\min})(1- \mathcal{E}_1^{\min}) =  1- \mathcal{E}_1^{\min}$ is the maximum attainable signaling fidelity. $0 \leq \mathcal{ F }^{\max} \leq 1$ is bounded for $\Delta G \geq 0$.

% ------------------------------- Figure -------------

\emph{Kinetics of resetting---}The landscape of the assembly supports both activation and deactivation (resetting) stages of the signaling cycle. As shown in Fig.~\ref{fig:landscape}a, activation proceeds predominantly through the  sequence $\sigma_{0}^{+} \to \sigma_{1}^{+} \to \sigma_{2}^{+}$. Deactivation proceeds predominantly through the  sequence $\sigma_{0}^{-} \to \sigma_{2}^{-} \to \sigma_{1}^{-}$. The reporter deactivates before the sensor deactivates, causing the assembly to switch as a last-in-first-out (LIFO) stack. 

\begin{figure}
\includegraphics[width=8.5cm]{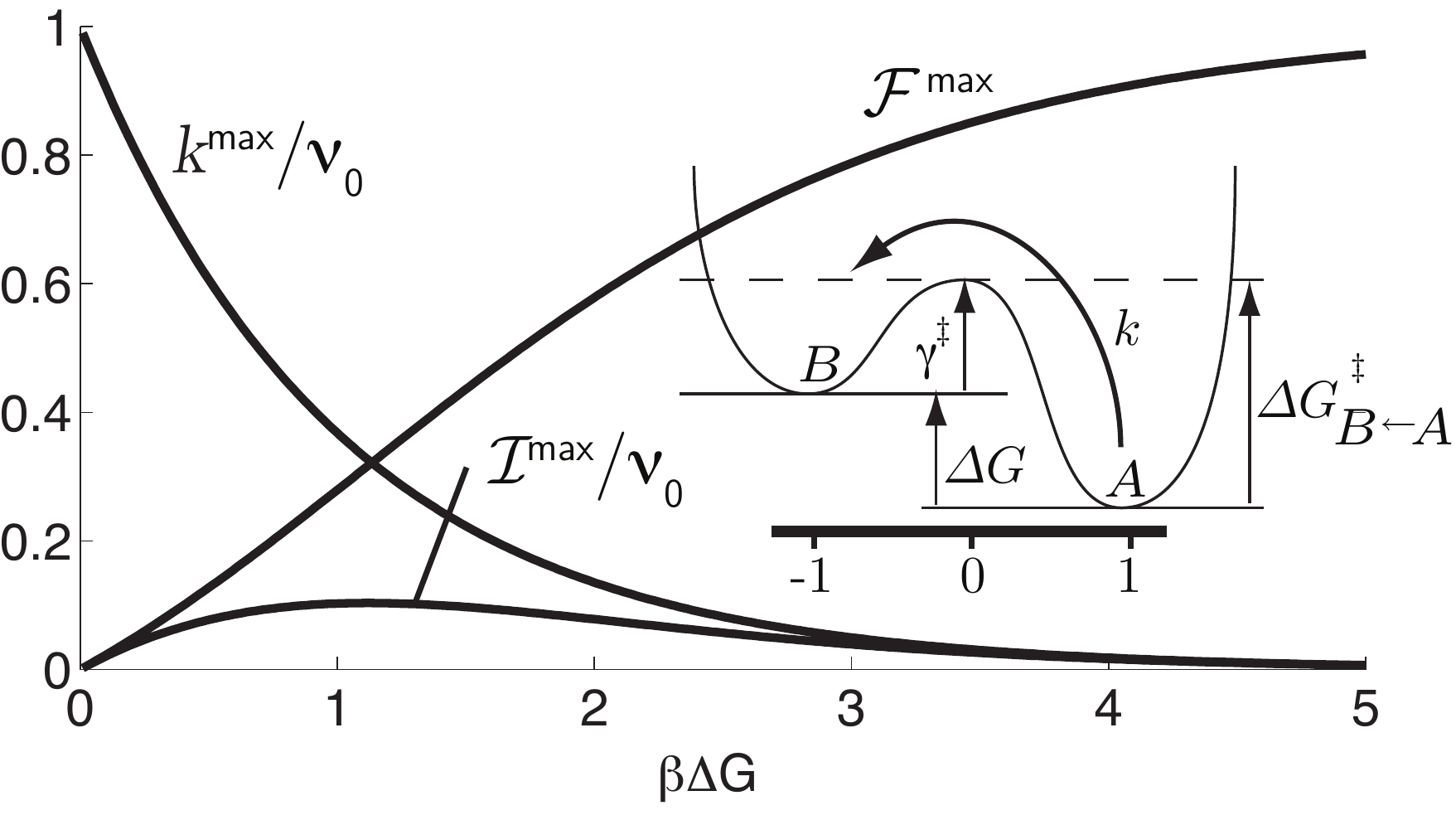}
\caption
{\label{fig:info_limit}
Trade-off between the deactivation rate $k^{\max}$ and signaling fidelity $\mathcal{F}^{\max}$ in rate of computation $\mathcal{I}^{\max}$ by an allosteric assembly. $\mathcal{I}^{\max} = k^{\max}\mathcal{F}^{\max}$. Inset: free energy surface for a network transition.
}
\end{figure}

While a large $\Delta G$ increases maximal activation and lowers $\mathcal{E}_1$, it also contributes to the free energy barrier (Fig~\ref{fig:info_limit}, inset) that must be overcome (transiently) for deactivation to proceed. The reporter must gain $-(\mathcal{D}_B - \mathcal{D}_A) = \beta \Delta G$ of information from the solvent before the sensor can deactivate. Here, $S_A = (0 \bs 11 )$, $S_B = (0 \bs 10 )$, and $\mathcal{D}_{j} = \mathcal{D}[ \delta(\set{S_i} - S_j) || \set{p_{i}} ]$~(\ref{eq:KL}). For a transition with free energy of activation $\Delta G^{\ddagger} = \Delta G + \gamma^{\ddagger}$ and barrier crossing attempt frequency $\nu$, the absolute transition rate $k_{B \from A} = \nu_0\Pi_{BA}$ satisfies the Arrhenius relation, $k = \nu \exp{\left[- \beta \Delta G^{\ddagger}  \right]} = \nu \exp{\left[\sigma^{\ddagger}/k_B  \right]}\exp{\left[- \beta \delta^{\ddagger}  \right]} =  \nu^{\prime} \exp{\left[- \beta \delta^{\ddagger}  \right]}$, where $\delta^{\ddagger}$ is the enthalpy of activation, $\sigma^{\ddagger}$ is the entropy of activation, and $\nu^{\prime}$ is the entropy of activation-adjusted barrier crossing attempt frequency. The rate of information gain ($\Delta G > 0$), $-\partial_{\tau} \mathcal{D} = - ( \mathcal{D}_{B} -  \mathcal{D}_{A}) \Pi_{BA}$ is bounded: $0 \leq -\partial_{\tau} \mathcal{D} \leq \beta \Delta G e^{- \beta \Delta G}$. 
%\begin{equation}
%\left\{ \begin{array}{rclr}
%0 \geq & -\partial_{\tau} \mathcal{D}  & \leq \beta \Delta G e^{- \beta \Delta G}, & \forall~\Delta G > 0 \\
%\beta \Delta G \geq & -\partial_{\tau} \mathcal{D}, & \leq 0 &\Delta G \leq 0
%\end{array} \right. .
%\label{eq:trade_off}
%\end{equation}
%[See EPAPS for derivation of (\ref{eq:trade_off})]. 
See EPAPS Document No. [x] for proof. The upper limit is approached as $\nu \to \nu_0$ and $\gamma^{\ddagger} \to 0$. $\max(-\partial_{\tau} \mathcal{D}) = e^{-1}$ at $\beta \Delta G = 1$, and $-\partial_{\tau} \mathcal{D} \to 0$ as $\Delta G \to \infty$. Therefore, a large $\Delta G$ reduces signaling error (\ref{eq:signaling_error}) but can critically slow deactivation. Indeed, temperature-dependent stopped-flow FRET measurements of the cardiac regulatory assembly~\cite{Robinson_2007b} show that deactivation of the reporter $(s_0 \bs 1 \sigma_2^{-})$ is the rate limiting step in deactivation with $\nu^{\prime} = 1.0 \times 10^{10}$ s$^{-1}$ and $\delta^{\ddagger} = 18.3~k_{B} T$ ($T = 15$ C). Lowering $\Delta G$ by 2 $k_B T$ would slow $(s_0 \bs 1 \sigma_2^{-})$ by 99\% ($k = 117 \to 1.0$ s$^{-1}$).

\emph{Physical limit---}The maximum rate of computation $\mathcal{I}^{max}$ (in bits/sec) is the maximum rate of the information-gaining step in deactivation $k^{\max} = \nu_0 \exp(-\beta \Delta G)$ times the maximum bit transmission fidelity $\mathcal{F}^{\max}$. We obtain (Fig~\ref{fig:info_limit})
\[
\mathcal{I}^{max} = \nu_0 e^{-\beta \Delta G} \left[1- \log_{2} \left( 1 + e^{- \beta \Delta G} \right) -  \frac{  \beta \Delta G e^{- \beta \Delta G}}{1 + e^{- \beta \Delta G}} \right],
\]
with $\max(\mathcal{I}^{max}) = 0.104 \nu_0$ at $\Delta G =  1.12 k_BT$. For a typical allosteric transition with wavenumber $\bar{\nu} = 10$ cm$^{-1}$, $\nu_0 = c \bar{\nu} = 3 \times 10^{11}$ \sec, and $\max(\mathcal{I}^{max}) = 3.1 \times 10^{10}$  bits/s. This is about 1000 times less than the Heisenberg limit on the rate of computation: $4 k_B T / h = 2.6 \times 10^{13}$ bits/s (T = 37 C)~\cite{Lloyd:2000lr}. The Heisenberg limit ignores system resetting.

\emph{Conclusions---}Cells employ assemblies of allosteric proteins---molecular switches that bind diffusing ligands and communicate binding as altered activity---to regulate intracellular function. These protein complexes are nano-scale Brownian computers whose function depends on random collisions with solvent. Elucidating the physical properties that limit the rate of Brownian computation is essential for understanding the molecular systems-level design of biological signaling complexes and the pathophysiology of diseases that involve these complexes.  The \Cm-sensitive cardiac TnC-TnI complex, an assembly of two allosteric proteins, is a paradigm for Brownian computation. The assembly functions as a digital logic buffer that transduces a binary input (the \Cm~bound/unbound status of the sensor TnC) to a  binary output (the  bi-metastable state of the reporter TnI). 

We examined the cardiac TnC-TnI complex using a non-equilibrium phenomenology of allostery in protein assemblies. The phenomenology assumes limited spatial extent of protein-protein interactions within a protein assembly, just as residue-residue interactions have limited spatial extend within a protein~\cite{Jacobs:2001gh, Jacobs:2003rv}. Computation occurs as a \Cm-induced perturbation of the the system's free energy landscape, causing a change in the population distribution among the set of coarse-grained system-states. Nearest-neighbor-limited influence, a spatial analog of the Markov assumption, constrains the topology of the system's free energy landscape and introduces degenerate transition probabilities. A single free energy change governs activation/deactivation of the reporter independent of whether \Cm~is bound to the sensor. Decreasing free energy drop enhances signaling fidelity but slows the rate of deactivation. The trade-off between fidelity and resetting speed physically limits the rate of information processing.  This physical limit is an emergent constraint~\cite{Landauer:1996hh, Davies:2004a} that arises from limited spatial extent of protein-protein interactions, a feature of the complex nature of the assembly. This limit is faced by assemblies that couple to a single chemical bath. Many assemblies appear to have addressed this limit by coupling to a second energy source. This energy source is usually inorganic phosphate derived from hydrolysis of nucleoside triphosphate. Examples include the family of G-protein coupled receptors, the NF-$\kappa$B family of transcription factors, and the troponin-tropomyosin-regulated interaction of actin and myosin in striated muscle.

%Allosteric signaling assemblies (allosteric assembly) are complex spatially extended systems that dissipatively transduce the free energy change conferred by ligand binding/release to a change the activity of the terminal component in the assembly. Nearest-neighbor-limited influence between the components creates degenerate transition probabilities for many system-state transitions. This degeneracy constrains the architecture of the allosteric assembly free energy landscape and imposes an upper limit on the rate of computation through a trade-off between the rate and reliability of computation. The relative importance of the two competing objectives---\emph{(i)} representing the ligand binding status at its output without error and \emph{(ii)} rapid deactivation (resetting)---is reflected in the architecture of the CRS free energy landscape, recovered from FRET measurements.

% ------------------------------- Figure -------------
%\begin{figure}
%%\centering
%%$\begin{array}{l}
%%\multicolumn{1}{l}{\mbox{\bf (a)}} \\ [-0.53cm] 
%%\includegraphics[width=8.5cm]{to5} \\
%%\multicolumn{1}{l}{\mbox{\bf (b)}} \\ %[-0.53cm] 
%%%\includegraphics[width=7cm]{landscape_both}
%%\includegraphics[width=8.5cm]{limits}
%%\end{array}$
%\includegraphics[width=8.5cm]{to5}

%\caption
%{\label{fig:trade_off}
%The rate of information gain $-\partial_{\tau} \mathcal{D}$ for an allosteric transition with net free energy change $\Delta G$. (\ref{eq:trade_off}) (gray area). }
%\end{figure}

 \section*{Acknowledgments}
%This work was supported in part by by the NIH (5R01HL052508-1, regulatory mechanisms) and by the European Union (MSCF-CT-2004-013119, Marie Curie Conference and Training Courses). I gratefully acknowledge the input and encouragement of Herbert Cheung, Ryoichi Kawai, and Ra\'ul Toral. 
This work was supported by the NIH and the European Union. I thank Herbert Cheung, Ryoichi Kawai, Ra\'ul Toral,  and Bart Cleuren for stimulating discussions and comments.

%The rate of information gain is a bounded function,
%\begin{equation}
%\left\{ \begin{array}{rclr}
%0 \geq &\partial_{\tau} \mathcal{D}  & \leq \beta G \exp(-  \beta G), & \forall~G > 0 \\
%G \geq & \partial_{\tau} \mathcal{D}, & \leq 0 & G \leq 0
%\end{array} \right. .
%\end{equation}
%The maximum possible rate of information gain occurs at the extremum $\partial_{ G} \beta G \exp(  \beta G) = 0$. Solving, one obtains $\max(\partial_{\tau} \mathcal{D} ) = e^{-1}$ at $\beta G = 1$ as the maximum possible rate of information gain.  There is no corresponding limit on the maximum rate of information loss.  

%$\partial_{\tau} \mathcal{D}$ should be interpreted as the rate of information gain per clock cycle.  It is a dimensionless quantity because time has been normalized against the clock speed, $\nu_0$, using $\tau := t \nu_0$.  

%\begin{equation*}
%%\frac{ \Pi_{j \from i} }{ \Pi_{j \to i} } = \exp ( -\beta  G_{j \from i} )
%\Pi_{j \from i} / \Pi_{j \to i} = \exp ( -\beta  G_{j \from i} ).
%\label{eq:detailed_balance}
%\end{equation*}

% Bibliography
%\bibliographystyle{prl}
%\bibliographystyle{plplain}
%\bibliography{/Users/jrobinson/Library/References/bibtex/refs_uni}

\begin{thebibliography}{23}
\expandafter\ifx\csname natexlab\endcsname\relax\def\natexlab#1{#1}\fi
\expandafter\ifx\csname bibnamefont\endcsname\relax
  \def\bibnamefont#1{#1}\fi
\expandafter\ifx\csname bibfnamefont\endcsname\relax
  \def\bibfnamefont#1{#1}\fi
\expandafter\ifx\csname citenamefont\endcsname\relax
  \def\citenamefont#1{#1}\fi
\expandafter\ifx\csname url\endcsname\relax
  \def\url#1{\texttt{#1}}\fi
\expandafter\ifx\csname urlprefix\endcsname\relax\def\urlprefix{URL }\fi
\providecommand{\bibinfo}[2]{#2}
\providecommand{\eprint}[2][]{\url{#2}}

\bibitem[{\citenamefont{Freire}(1999)}]{Freire:1999fk}
\bibinfo{author}{\bibfnamefont{E.}~\bibnamefont{Freire}},
  \bibinfo{journal}{Proc Natl Acad Sci U S A} \textbf{\bibinfo{volume}{96}},
  \bibinfo{pages}{10118} (\bibinfo{year}{1999}).

\bibitem[{\citenamefont{Hilser et~al.}(2006)\citenamefont{Hilser,
  Garcia-Moreno~E, Oas, Kapp, and Whitten}}]{Hilser:2006pb}
\bibinfo{author}{\bibfnamefont{V.~J.} \bibnamefont{Hilser}},
  \bibinfo{author}{\bibfnamefont{B.}~\bibnamefont{Garcia-Moreno~E}},
  \bibinfo{author}{\bibfnamefont{T.~G.} \bibnamefont{Oas}},
  \bibinfo{author}{\bibfnamefont{G.}~\bibnamefont{Kapp}}, \bibnamefont{and}
  \bibinfo{author}{\bibfnamefont{S.~T.} \bibnamefont{Whitten}},
  \bibinfo{journal}{Chem Rev} \textbf{\bibinfo{volume}{106}},
  \bibinfo{pages}{1545} (\bibinfo{year}{2006}).

\bibitem[{\citenamefont{Kern and Zuiderweg}(2003)}]{Kern:2003uk}
\bibinfo{author}{\bibfnamefont{D.}~\bibnamefont{Kern}} \bibnamefont{and}
  \bibinfo{author}{\bibfnamefont{E.~R.~P.} \bibnamefont{Zuiderweg}},
  \bibinfo{journal}{Curr Opin Struct Biol} \textbf{\bibinfo{volume}{13}},
  \bibinfo{pages}{748} (\bibinfo{year}{2003}).

\bibitem[{\citenamefont{Hartwell et~al.}(1999)\citenamefont{Hartwell, Hopfield,
  Leibler, and Murray}}]{Hartwell:1999fk}
\bibinfo{author}{\bibfnamefont{L.~H.} \bibnamefont{Hartwell}},
  \bibinfo{author}{\bibfnamefont{J.~J.} \bibnamefont{Hopfield}},
  \bibinfo{author}{\bibfnamefont{S.}~\bibnamefont{Leibler}}, \bibnamefont{and}
  \bibinfo{author}{\bibfnamefont{A.~W.} \bibnamefont{Murray}},
  \bibinfo{journal}{Nature} \textbf{\bibinfo{volume}{402}},
  \bibinfo{pages}{C47} (\bibinfo{year}{1999}).

\bibitem[{\citenamefont{Pawson and Nash}(2003)}]{Pawson:2003oz}
\bibinfo{author}{\bibfnamefont{T.}~\bibnamefont{Pawson}} \bibnamefont{and}
  \bibinfo{author}{\bibfnamefont{P.}~\bibnamefont{Nash}},
  \bibinfo{journal}{Science} \textbf{\bibinfo{volume}{300}},
  \bibinfo{pages}{445} (\bibinfo{year}{2003}).

\bibitem[{\citenamefont{Volkman et~al.}(2001)\citenamefont{Volkman, Lipson,
  Wemmer, and Kern}}]{Volkman:2001bs}
\bibinfo{author}{\bibfnamefont{B.~F.} \bibnamefont{Volkman}},
  \bibinfo{author}{\bibfnamefont{D.}~\bibnamefont{Lipson}},
  \bibinfo{author}{\bibfnamefont{D.~E.} \bibnamefont{Wemmer}},
  \bibnamefont{and} \bibinfo{author}{\bibfnamefont{D.}~\bibnamefont{Kern}},
  \bibinfo{journal}{Science} \textbf{\bibinfo{volume}{291}},
  \bibinfo{pages}{2429} (\bibinfo{year}{2001}).

\bibitem[{\citenamefont{Voet and Voet}(2004)}]{Voet:2004a}
\bibinfo{author}{\bibfnamefont{D.}~\bibnamefont{Voet}} \bibnamefont{and}
  \bibinfo{author}{\bibfnamefont{J.~G.} \bibnamefont{Voet}},
  \emph{\bibinfo{title}{Biochemistry}} (\bibinfo{publisher}{Wiley, New York},
  \bibinfo{year}{2004}), \bibinfo{edition}{3rd} ed.

\bibitem[{\citenamefont{Bennett}(1982)}]{Bennett:1982aa}
\bibinfo{author}{\bibfnamefont{C.~H.} \bibnamefont{Bennett}},
  \bibinfo{journal}{Int J Theor Phys} \textbf{\bibinfo{volume}{21}},
  \bibinfo{pages}{905} (\bibinfo{year}{1982}).

\bibitem[{\citenamefont{Leff and Rex}(2003)}]{Rex:2003ne}
\bibinfo{author}{\bibfnamefont{H.~S.} \bibnamefont{Leff}} \bibnamefont{and}
  \bibinfo{author}{\bibfnamefont{A.~F.} \bibnamefont{Rex}},
  \emph{\bibinfo{title}{Maxwell's Demon 2}} (\bibinfo{publisher}{IOP,
  Philidelphia}, \bibinfo{year}{2003}), \bibinfo{edition}{2nd} ed.

\bibitem[{\citenamefont{Szilard}(1929)}]{Szilard:1929a}
\bibinfo{author}{\bibfnamefont{L.}~\bibnamefont{Szilard}}, \bibinfo{journal}{Z
  Phys} \textbf{\bibinfo{volume}{53}}, \bibinfo{pages}{840}
  (\bibinfo{year}{1929}).

\bibitem[{\citenamefont{Landauer}(1961)}]{Landauer:1961a}
\bibinfo{author}{\bibfnamefont{R.}~\bibnamefont{Landauer}},
  \bibinfo{journal}{IBM J Res Dev} \textbf{\bibinfo{volume}{5}},
  \bibinfo{pages}{183} (\bibinfo{year}{1961}).

\bibitem[{\citenamefont{Norton}(2005)}]{Norton:2005lr}
\bibinfo{author}{\bibfnamefont{J.~D.} \bibnamefont{Norton}},
  \bibinfo{journal}{Stud Hist Phil Sci, B} \textbf{\bibinfo{volume}{36}},
  \bibinfo{pages}{375} (\bibinfo{year}{2005}).

\bibitem[{\citenamefont{Lloyd}(2000)}]{Lloyd:2000lr}
\bibinfo{author}{\bibfnamefont{S.}~\bibnamefont{Lloyd}},
  \bibinfo{journal}{Nature} \textbf{\bibinfo{volume}{406}},
  \bibinfo{pages}{1047} (\bibinfo{year}{2000}).

\bibitem[{\citenamefont{Kobayashi and Solaro}(2005)}]{Kobayashi:2005fk}
\bibinfo{author}{\bibfnamefont{T.}~\bibnamefont{Kobayashi}} \bibnamefont{and}
  \bibinfo{author}{\bibfnamefont{R.~J.} \bibnamefont{Solaro}},
  \bibinfo{journal}{Annu Rev Physiol} \textbf{\bibinfo{volume}{67}},
  \bibinfo{pages}{39} (\bibinfo{year}{2005}).

\bibitem[{\citenamefont{H\"anggi et~al.}(1990)\citenamefont{H\"anggi, Talkner,
  and Borkovec}}]{Hanggi_1990a}
\bibinfo{author}{\bibfnamefont{P.}~\bibnamefont{H\"anggi}},
  \bibinfo{author}{\bibfnamefont{P.}~\bibnamefont{Talkner}}, \bibnamefont{and}
  \bibinfo{author}{\bibfnamefont{M.}~\bibnamefont{Borkovec}},
  \bibinfo{journal}{Rev Mod Phys} \textbf{\bibinfo{volume}{62}},
  \bibinfo{pages}{251} (\bibinfo{year}{1990}).

\bibitem[{\citenamefont{Berezhkovskii and Szabo}(2005)}]{Berezhkovskii:2005lr}
\bibinfo{author}{\bibfnamefont{A.}~\bibnamefont{Berezhkovskii}}
  \bibnamefont{and} \bibinfo{author}{\bibfnamefont{A.}~\bibnamefont{Szabo}},
  \bibinfo{journal}{J Chem Phys} \textbf{\bibinfo{volume}{122}},
  \bibinfo{pages}{014503} (\bibinfo{year}{2005}).

\bibitem[{\citenamefont{Bahar et~al.}(1998)\citenamefont{Bahar, Atilgan,
  Demirel, and Erman}}]{Bahar:1998a}
\bibinfo{author}{\bibfnamefont{I.}~\bibnamefont{Bahar}},
  \bibinfo{author}{\bibfnamefont{A.~R.} \bibnamefont{Atilgan}},
  \bibinfo{author}{\bibfnamefont{M.~C.} \bibnamefont{Demirel}},
  \bibnamefont{and} \bibinfo{author}{\bibfnamefont{B.}~\bibnamefont{Erman}},
  \bibinfo{journal}{Phys Rev Lett} \textbf{\bibinfo{volume}{80}},
  \bibinfo{pages}{2733} (\bibinfo{year}{1998}).

\bibitem[{\citenamefont{Robinson et~al.}(2008)\citenamefont{Robinson, Dong, and
  Cheung}}]{Robinson_2007b}
\bibinfo{author}{\bibfnamefont{J.~M.} \bibnamefont{Robinson}},
  \bibinfo{author}{\bibfnamefont{W.-J.} \bibnamefont{Dong}}, \bibnamefont{and}
  \bibinfo{author}{\bibfnamefont{H.~C.} \bibnamefont{Cheung}},
  \bibinfo{journal}{Biophys J, submitted}  (\bibinfo{year}{2008}).

\bibitem[{\citenamefont{Cover and Thomas}(2006)}]{Cover_2006a}
\bibinfo{author}{\bibfnamefont{T.~M.} \bibnamefont{Cover}} \bibnamefont{and}
  \bibinfo{author}{\bibfnamefont{J.~A.} \bibnamefont{Thomas}},
  \emph{\bibinfo{title}{Elements of Information Theory}}
  (\bibinfo{publisher}{Wiley, Hoboken, New Jersey}, \bibinfo{year}{2006}),
  \bibinfo{edition}{2nd} ed.

\bibitem[{\citenamefont{Jacobs et~al.}(2001)\citenamefont{Jacobs, Rader, Kuhn,
  and Thorpe}}]{Jacobs:2001gh}
\bibinfo{author}{\bibfnamefont{D.~J.} \bibnamefont{Jacobs}},
  \bibinfo{author}{\bibfnamefont{A.~J.} \bibnamefont{Rader}},
  \bibinfo{author}{\bibfnamefont{L.~A.} \bibnamefont{Kuhn}}, \bibnamefont{and}
  \bibinfo{author}{\bibfnamefont{M.~F.} \bibnamefont{Thorpe}},
  \bibinfo{journal}{Proteins} \textbf{\bibinfo{volume}{44}},
  \bibinfo{pages}{150} (\bibinfo{year}{2001}).

\bibitem[{\citenamefont{Jacobs et~al.}(2003)\citenamefont{Jacobs, Dallakyan,
  Wood, and Heckathorne}}]{Jacobs:2003rv}
\bibinfo{author}{\bibfnamefont{D.~J.} \bibnamefont{Jacobs}},
  \bibinfo{author}{\bibfnamefont{S.}~\bibnamefont{Dallakyan}},
  \bibinfo{author}{\bibfnamefont{G.~G.} \bibnamefont{Wood}}, \bibnamefont{and}
  \bibinfo{author}{\bibfnamefont{A.}~\bibnamefont{Heckathorne}},
  \bibinfo{journal}{Physical Review E} \textbf{\bibinfo{volume}{68}}
  (\bibinfo{year}{2003}).

\bibitem[{\citenamefont{Landauer}(1996)}]{Landauer:1996hh}
\bibinfo{author}{\bibfnamefont{R.}~\bibnamefont{Landauer}},
  \bibinfo{journal}{Phys Lett A} \textbf{\bibinfo{volume}{217}},
  \bibinfo{pages}{188} (\bibinfo{year}{1996}).

\bibitem[{\citenamefont{Davies}(2004)}]{Davies:2004a}
\bibinfo{author}{\bibfnamefont{P.~C.~W.} \bibnamefont{Davies}},
  \bibinfo{journal}{Complexity} \textbf{\bibinfo{volume}{10}},
  \bibinfo{pages}{11} (\bibinfo{year}{2004}).

\end{thebibliography}

\begin{thebibliography}{1}
\expandafter\ifx\csname natexlab\endcsname\relax\def\natexlab#1{#1}\fi
\expandafter\ifx\csname bibnamefont\endcsname\relax
  \def\bibnamefont#1{#1}\fi
\expandafter\ifx\csname bibfnamefont\endcsname\relax
  \def\bibfnamefont#1{#1}\fi
\expandafter\ifx\csname citenamefont\endcsname\relax
  \def\citenamefont#1{#1}\fi
\expandafter\ifx\csname url\endcsname\relax
  \def\url#1{\texttt{#1}}\fi
\expandafter\ifx\csname urlprefix\endcsname\relax\def\urlprefix{URL }\fi
\providecommand{\bibinfo}[2]{#2}
\providecommand{\eprint}[2][]{\url{#2}}

\bibitem[{\citenamefont{H\"anggi et~al.}(1990)\citenamefont{H\"anggi, Talkner,
  and Borkovec}}]{Hanggi_1990a}
\bibinfo{author}{\bibfnamefont{P.}~\bibnamefont{H\"anggi}},
  \bibinfo{author}{\bibfnamefont{P.}~\bibnamefont{Talkner}}, \bibnamefont{and}
  \bibinfo{author}{\bibfnamefont{M.}~\bibnamefont{Borkovec}},
  \bibinfo{journal}{Rev Mod Phys} \textbf{\bibinfo{volume}{62}},
  \bibinfo{pages}{251} (\bibinfo{year}{1990}).

\end{thebibliography}

\clearpage
\appendix

\section{EPAPS}
Individual Markov transition probabilities $\Pi_{j \from i}$ depend on the barrier crossing free energies $\Delta G_{j \from i}^{\ddag}$ through the Arrhenius relation~\cite{Hanggi_1990a},
\begin{equation}
\Pi_{j \from i} = \left\{ \begin{array}{cr}
N_{i} \widehat{ \nu_{j \from i} } \exp ( -\beta  \Delta G_{j \from i}^{\ddag} ), & \forall \, j \neq i \\
N_{i} \left(1 - \sum_{j  \neq i} \Pi_{j \from i} \right), & j = i
\end{array} \right. ,
\label{eq:Arrhenius}
\end{equation}
with normalized barrier crossing attempt frequencies $\widehat{ \nu_{j \from i} } =  \nu_{j \from i} /  \nu_0$, $\nu_0 = \max( \set{ \nu_{j \from i} } )$. $\beta = 1/ k_{\mathrm{B}} T$. The normalization constants $N_{i} = \sum_{j} \Pi_{j \from i} $ ensure that $\sum_{j} \Pi_{j \from i} = 1$, as required. All transition probabilities are bounded, $0 \leq \Pi_{j \from i} \leq 1$.  The barrier crossing free energies are calcuated using
\begin{equation}
\Delta G_{j \from i}^{\ddag} = \left\{ \begin{array}{cr}
\gamma^{\ddag} + \Delta G, &  \forall~ \Delta G > 0 \\
\gamma^{\ddag},  &  \Delta G \leq 0 
\end{array} \right.
\label{eq:G_dagger}
\end{equation}
where $\gamma^{\ddag} \geq 0$ and $\Delta G = G_{j} - G_{i}$. Furthermore, we set $N_{i} \widehat{ \nu_{j \from i} } = N_{j} \widehat{ \nu_{j \to i} }$ to satisfy microscopic balance, $\Pi_{j \from i} / \Pi_{j \to i} = \exp ( -\beta  G_{j \from i} )$.

Substituting (\ref{eq:G_dagger}) into (\ref{eq:Arrhenius}) we obtain
\begin{equation}
\Pi_{j \from i} =  N_{i} \widehat{ \nu_{j \from i} } 
\left\{ \begin{array}{lr}
\exp( - \beta ( \gamma^{\ddag} + \Delta G)),  & \forall~ \Delta G > 0 \\
\exp( - \beta \gamma^{\ddag}), & \Delta G \leq 0 
\end{array} \right.,
\label{eq:ineq}
\end{equation}
with the constraints $0 \geq N_{i}, \widehat{ \nu_{j \from i} } \leq 1$ and $\gamma^{\ddag} \geq 0$. Effective barrier crossing attempt frequencies (s$^{-1}$) are defined $\nu_{j \from i} = \nu_0 N_{i} \widehat{ \nu_{j \from i} }$. %Absolute rate constants are defined $k_{j \from i} = \nu_{0} \Pi_{j \from i}$.

\section*{Proof of the equation for the bounded rate of information gain}
%\appendix

The relative entropy change for the transition $\mathbf{ p}_0 \to \mathbf{ p}$ from an initial distribution $\mathbf{ p}_0$ to a second distribution $\mathbf{ p}$ is
\[
\mathcal{D}[ \mathbf{ p}_0 || \mathbf{ p}] = \sum_{\Omega} \mathbf{ p}_0 \ln[\mathbf{ p}_0/ \mathbf{ p}].
\]
Systems will relax to the equilibrium density $\mathbf{ p}^*$ given by the Boltzmann distribution
\begin{equation}
p^*_i(\mu) = \frac{1}{Z} e^{-\beta G(S_i; \mu) },
\label{eq:Boltz_dist}
\end{equation}
where $Z = \sum_{\Omega} \exp{-\beta G(S_i; \mu) }$ is the partition function. Taking the system from state $S_A = (0 \bs 11)$ with $\mathbf{ p} = \delta(S - S_A)$ to its equilibrium distribution $\mathbf{ p}^*$, involves an information loss of $\mathcal{D}_{A} = \mathcal{D}[ \mathbf{ p} || \mathbf{ p}^*] = -\ln p_A$. Similarly taking the system from state $S_B = (0 \bs 11)$ with $\mathbf{ p} = \delta(S - S_B)$ to its equilibrium distribution $\mathbf{ p}^*$, involves an information loss of $\mathcal{D}_{B} = \mathcal{D}[ \mathbf{ p} || \mathbf{ p}^*] = -\ln p_B$. The net \emph{gain} of information for $S_A \to S_B$ is $-(\mathcal{D}_{B} - \mathcal{D}_{A}) = \ln[p_B/ p_A]$.  Inserting (\ref{eq:Boltz_dist}), rearranging and canceling $Z$, the net gain of information is
\begin{equation}
-(\mathcal{D}_{B} - \mathcal{D}_{A}) = \beta \Delta G,
\label{eq:info_gain}
\end{equation}
where $\Delta G = G(S_B) - G(S_A)$.  When $\Delta G > 0$, there is a gain of information.

\begin{figure}[t]
\includegraphics[width=8.5cm]{\FIGS/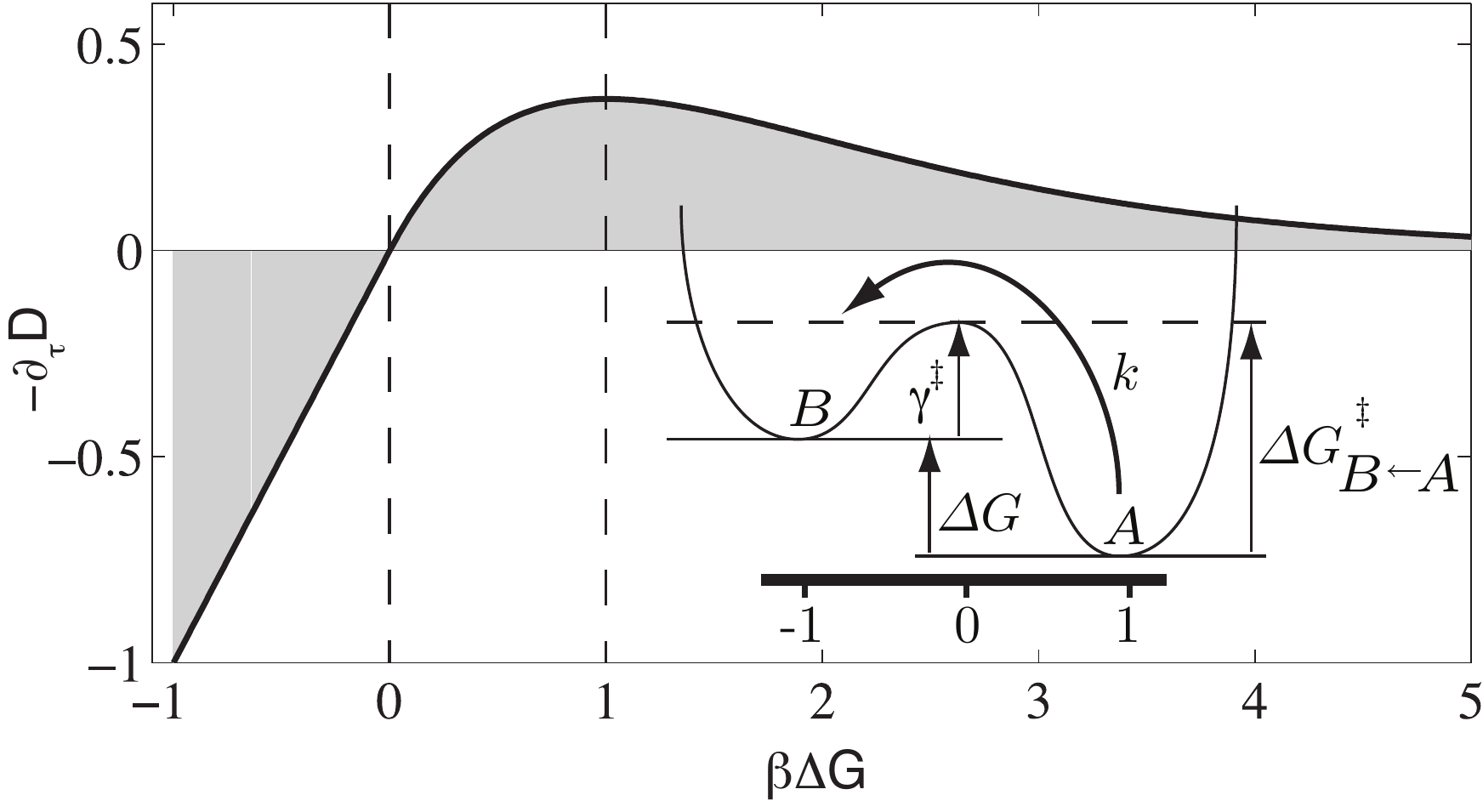}
\caption
{\label{fig:trade_off}
The rate of information gain $-\partial_{\tau} \mathcal{D}$ for an allosteric transition with net free energy change $\Delta G$. (\ref{eq:gain_rate}) (gray area). }
\end{figure}

The rate of information gain for $S_A \to S_B$ is defined, $-\partial_{\tau} \mathcal{D} = -( \mathcal{D}_{B} -  \mathcal{D}_{A}) \Pi_{B \from A}$. Substituting (\ref{eq:info_gain}) and (\ref{eq:Arrhenius}) we obtain
\[
-\partial_{\tau} \mathcal{D} = \beta \Delta G N_{A} \widehat{ \nu_{B \from A} } \exp( - \beta \Delta  G^{\ddag}_{B \from A}).
\]
Finally, substituting for $\Delta G^{\ddag}$ (\ref{eq:G_dagger}), we obtain
\[
-\partial_{\tau} \mathcal{D} = \beta G N_{A} \widehat{ \nu_{B \from A} } 
\left\{ \begin{array}{lr}
\exp( - \beta ( \gamma^{\ddag} + \Delta G)),  & \forall~\Delta G > 0 \\
\exp( - \beta \gamma^{\ddag}), & \Delta G \leq 0 
\end{array} \right..
\]
But $N_{A} \leq 0$, $\widehat{ \nu_{B \from A} } \leq 1$ and $\gamma^{\ddag} \geq 0$. The rate of information gain is a bounded function (Fig.~\ref{fig:trade_off}),
%\begin{equation}
%\left\{ \begin{array}{rclr}
%0 \geq &\partial_{\tau} \mathcal{D}  & \leq \beta G \exp(-  \beta G), & \forall~G > 0 \\
%G \geq & \partial_{\tau} \mathcal{D}, & \leq 0 & G \leq 0
%\end{array} \right. .
%\end{equation}
\begin{equation}
\left\{ \begin{array}{rclr}
0 \leq & -\partial_{\tau} \mathcal{D}  & \leq \beta \Delta G \exp( - \beta \Delta G), & \forall~\Delta G > 0 \\
\beta \Delta G \leq & -\partial_{\tau} \mathcal{D}, & \leq 0 &\Delta G \leq 0
\end{array} \right.
\label{eq:gain_rate}
\end{equation}
The maximum possible rate of information gain occurs at the extremum ${\partial(-\partial_{\tau} \mathcal{D} )}{/\partial {\Delta G}}  = 0$. Solving, we obtain $\max(-\partial_{\tau} \mathcal{D} ) = e^{-1}$ at $\beta \Delta G = 1$ as the maximum possible rate of information gain.  There is no corresponding limit on the maximum rate of information loss.  

$-\partial_{\tau} \mathcal{D}$ should be interpreted as the rate of information gain per clock cycle.  It is a dimensionless quantity because time has been normalized against the clock speed $\nu_0$ using $\tau \equiv t \nu_0$.  

%%-----------------------------------------------------------------------------------------------------------------------------
\end{document}